# Ride N' Rhythm, Bike as an Embodied Musical Instrument to Improve Music Perception for Young Children

**Weina Jin**
School of Interactive Arts and Technology
Simon Fraser University
Surrey, BC, Canada
weinaj@sfu.ca

**Diane Gromala**
School of Interactive Arts and Technology
Simon Fraser University
Surrey, BC, Canada
gromala@sfu.ca

**Alissa N. Antle**
School of Interactive Arts and Technology
Simon Fraser University
Surrey, BC, Canada
aantle@sfu.ca

## Abstract
Music plays a crucial role in young children's development. Current research lacks the design of an interactive system for younger children that could generate dynamic music change in response to the children's body movement. In this paper, we present the design of bike as an embodied musical instrument for young children 2-5 years old to improve their music perception skills. In the *Ride N' Rhythm* prototype, the rider's body position maps to the music volume; and the speed of the bike maps to the tempo. The design of the prototype incorporates the Embodied Music Cognition theory and Dalcroze Eurhythmics pedagogy, and aims to internalize the "intuitive" knowing and musical understanding via the combination of music and body movement.

## ACM Classification Keywords
H.5.2. Information interfaces and presentation (e.g., HCI): User Interfaces; K.3.0 Computers and Education: General.

**Introduction**
Music plays an essential role in early childhood development. For toddlers and preschoolers, the systematic exposure to music improves their spatial and temporal reasoning [13], support the development of emotional state and concentration [7]. Music perception skills are also correlated with the improvement of language development [5], and mathematics skills [17].

The experience of body movement plays an important role in musical perception even at the infant stage, as evidenced in neuroscience studies [12]. Other experiments also suggested that body movements performed to music will have an influence on the perception of musical features, such as rhythm and pitch [11]. Such research fall in the paradigm of Embodied Music Cognition, which claims that music cognition is strongly determined by bodily mediated interactions with music [10]. As to the practice field of music education, Dalcroze suggested that internalized musical understanding should be based on the combination of music and bodily experiences [8]. His pedagogy of applying body movement in teaching music is known as Dalcroze Eurhythmics and has been widely adopted by music teachers. According to Dalcroze Eurhythmics, the body movement bridges between concrete musical activities and abstract conceptual thinking by acting as an embodied metaphor. Concretely, the abstract musical qualities such as high/low, or fast/slow can be reflected by analogous body movement embodying these concepts and thus be understood in the most primal way without the need to name them [9]. The pre-reflective bodily knowing is often regarded as unconscious or intuitive [1]. Later, such "intuitive" knowing will be extracted when the child is taught the conceptual form of music such as naming and notation.

For younger children 2-5 years old, the current early childhood music education curriculums that are in align with the Dalcroze Eurhythmics are usually designed to instruct children to move or dance to the music [6]. There lacks an interactive system that could create dynamic music in response to children's body movement. Inspired by the above theory and evidence, in this research, we aim to design and build a tangible user interface (TUI) for young children 2-5 years old. We first examine the previous work on embodied TUI for children's music learning, and identify the research gap. Then we describe the design rationale and the prototype structure. We also propose two usability studies for validation.

**Related Work**
A growing body of work has focused on designing TUIs that addressed the embodied interaction and involved body movement for children to learn music. For example, Zigelbaum et al. designed a full-body musical interfaces *BodyBeats* to encourage physical play and foster pattern learning of children [16]. Volpe, et al., described *BeSound* that allow children to explore rhythm, melody, and harmony by whole-body movement [14]. *MoSo* is a set of physical interactive TUIs that allow children 7-9 years old to manipulate the pitch, volume and tempo of ongoing tones, in order to structure their understanding of these abstract sound concepts [4]. Besides the exploration of a variety of TUI applications, other research provided implications and suggestions for the future design of TUIs by conducting usability studies. Antle et al. examined the body-based metaphor mapping vs. non-metaphor mapping for sound outputs, and found that with the embodied sound metaphor mapping, children aged 7 to 10 were able to demonstrate sound sequences more accurately but resorted to spatial rather than body-based metaphors; also, the mapping must be easily discoverable as well as metaphorical to provide benefit for learning [2]. Bakker et al. identified and categorized commonly used metaphors based on 65 children's enactments of changing sound concepts. For example, the sound volume is related to the embodied schema of low/high, whereas the tempo is related to slow/fast speed [3].

Despite the recent work, few research has addressed the problem of designing TUIs for younger children 2-5 years old. In this research, we aim to design and build a TUI for toddlers and preschoolers to gain music perception skills via embodied interaction.

**Design Rationale**
**Structured vs. Unstructured Movement** Some of the previous research focused on unstructured body movement, which the TUI system support improvised body movement such as waving, clapping, rotating, jumping, stepping, etc. While this approach is feasible for school-aged children, younger children may find it difficult to control the sound parameters by unstructured movement. In addition, because of the existence of multiple metaphors [3], the mapping between the unstructured movement and the sound parameter change will cause confusion to younger children, and thus may not easily understand the sound change the based on the embodied metaphor. Therefore, to design a TUI that purposely improves young children's music perception skills, we employ the structured movement in our design.

**Whole-Body vs. Partial-Body Interaction** With the structured movement as the design requirement, it is more obvious to design an interactive form that involves only the upper or lower body movement, rather than the whole body. However, since feeling the music through the whole body with kinesthetic sensations rather than solely the ears will allow the children be more capable of recognizing the musical qualities [9], we decided to employ the whole body interaction. It may also have the additional effect of encouraging more physical activities and avoiding the sedentary lifestyle.

Since the whole-body movement must be structured, we are left with limited options. Cycling is the most viable solution and a very popular activity for young children 2-5 years old. When riding a bike, a child is experiencing a structured and rhythmic movement with his/her legs while the hands are responsible for fine-tuning the direction and posture. This mimics the play of many musical instruments, such as violin, cello, or guitar, where one needs to generate harmonious music by rhythmic gross movement, and create changes in volume or pitch by unnoticeable movement.

**Sound Parameters** The change of the sound parameters, such as volume, tempo, pitch, must be well-perceivable by younger children, and the mapping to the biking movement should be intuitive and in accordance with the embodied metaphor. Based on [3], we included only two sound parameters which are the most understandable ones: tempo and volume. The slow/fast tempo is controlled by the slow/fast speed of the bike, and the high/low volume is controlled by the upright/forward body position on the bike.

**The *Ride N' Rhythm* Prototype**
To examine our proposed design, we built the *Ride N' Rhythm* prototype. It turns a kids' bike into an embodied musical instrument. To capture the speed of the bike, we equipped the bike with a 3-axis accelerometer sensor; to detect the rider's body position, we equipped an ultrasonic sensor and on the bike frame facing the rider. We used a speaker to cycle through a piece of drum beats, with the tempo controlled by the accelerometer reading, and the volume controlled by the ultrasonic sensor reading. The sensors and speaks were connected via Arduino board to a Processing 3.0 program. The equipment is bundled together and attached to the front frame of the bike firmly to ensure it poses no hazards during the riding (Figure 1).

As soon as a child starts to ride the *Ride N' Rhythm* bike, s/he will hear the rhythmic beats. As s/he gradually speeds up, the tempo of the beats increases, and decreases when the child slows down. When the child leans forward to lower his/her body, the sound volume decreases; and it increases when the child sits upright. The music play stops when the ride ceases.

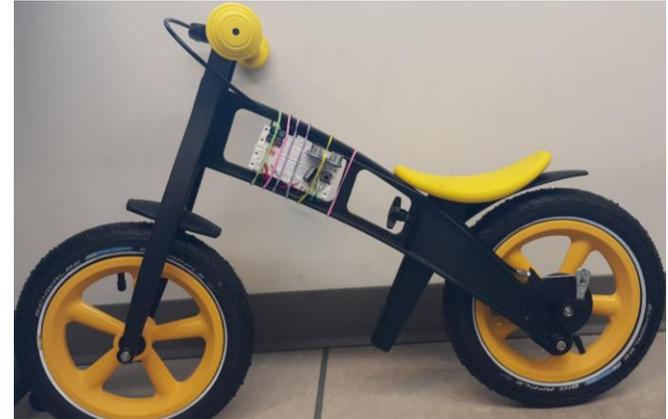

**Figure 1.** The *Ride N' Rhythm* prototype. It is equipped with an ultrasonic sensor to detect the body position; and an accelerometer to detect the speed.

The prototype consists of 3 modes. In mode 1, only the volume of the music can be changed by raising or lowering the body position; in mode 2, only the tempo of the music can be changed by speeding up or slowing down the bike; in mode 3, both the volume and the tempo can be changed. The parent can press the button on the prototype to switch among the modes, so that the sound change separately with tempo or volume. This design is to ease the cognitive load for young children. When the child gets familiar with the change of a single sound parameter, the mode 3 provides a comprehensive mode to control the tempo and volume simultaneously.

**Validation Plan**
We present a validation plan to evaluate the prototype. The general research question is: With the bike as a musical instrument where the tempo of the music is controlled by the riding speed, and the volume is controlled by the body position, will young children 2-5 years old gain a better music perception skills regarding volume and tempo?

The validation plan includes two studies: the first study is a proof-of-concept usability study with a qualitative study design. It will examine the following outcomes:

1) Whether the child can perceive the sound parameter changes while they are riding and changing speed and body posture, after a period of usage of the prototype;

2) Whether the child could use the bike prototype to purposely create the desired sound change, after a period of usage of the prototype.

Children who are 2-5 years old, are proficient in riding a balance bike, and can following basic verbal instructions will be enrolled in this study. The participants will ride the bike prototype for 20 minutes, and then a short interview will be conducted to ask the participant if they notice the sound changes during riding, if so, did they discover which action causes the sound changes. The researcher will teach the participant if s/he didn't find out how to control the sound the by their own. If a participant passes the above steps, s/he will be asked to create a fast/slow tempo or a low/high volume sound by riding the bike. The participant's performance and the total learning time will be recorded. The prototype testing and the interview will be video-recorded for qualitative analysis.

Based on the pilot study, the second study is a quasi-experiment design that evaluates whether the prototype will improve young children's music perception skills. Children who are 2-5 years old, are proficient in riding a balance bike, and can following basic verbal instructions will be enrolled in this study. Before enrolment, the parents will be informed that children are not allowed to attend any music classes during the study period.

After obtaining the consent form from the child's guardian, a pretest will be conducted to test the participant's perception and understanding of tempo and volume. The participant will be present with several pieces of music and be asked to select the one with high/low volume, or fast/slow tempo. Then, the participant will use the prototype for 20 minutes per day for three days. Lastly, the participant will receive a posttest which is the same as the pretest. We will run a paired t-test to analyze if there is statistically significant change after the intervention. Depending on the participant's verbal communication ability, a short interview may also be conducted regarding children's attitude towards the system.

**Conclusions**

In this paper, we present the design and development of the *Ride N' Rhythm* prototype based on Embodied Music Cognition and Dalcroze Eurhythmics. It extends a kid's bike to an embodied musical instrument by allowing the child to control the volume of the music with the body position, and to control the tempo with the riding speed. The dynamic change of the music following the child's cycling movement aims to create the "intuitive" knowing that improves young children's music perception skills. Additionally, it may also improve motor skills [15], and encourages more outdoor physical activities by creating more fun during cycling. Our work contributes to the music education and TUI community by providing a working prototype that facilitates early childhood development in music perception. The future work includes the evaluation the prototype on the improvement on young children's music perception skills.